\documentstyle[12pt]{article}
\begin{document}



\vskip 2truecm

\begin{center}

\vskip 2truecm

{\Large \bf
Solutions in Self-dual Gravity}
{\Large \bf Constructed
Via Chiral Equations}\\[8mm]

{ H. Garc\' \i a-Compe\'an and Tonatiuh Matos}

\vskip 1truecm
{\it Departamento de F\'{\i}sica}

{\it Centro de Investigaci\'on y de Estudios Avanzados del I.P.N.\\
Apdo. Post. 14-740, 07000, M\'exico, D.F., M\'exico.}

\end{center}

\begin{abstract}
The chiral model for self-dual gravity given by Husain in the context of
the chiral equations approach is discussed.  A Lie
algebra corresponding to a finite dimensional subgroup of the group of
symplectic diffeomorphisms is found, and then
use for expanding the Lie algebra valued connections associated with
the chiral model. The self-dual metric can be explicitly
given in terms of harmonic maps and in terms of a basis of this subalgebra.
\end{abstract}


\newpage
\begin{section}
{Introduction}
Since the introduction of Ashtekar's variables in General Relativity
\cite{ash} they were quickly applied to self-dual gravity. Later,
Ashtekar, Jacobson and Smolin (AJS) considered a new formulation of
half-flat solutions to Einstein's equations. To be more precise, making a
decomposition of a real 4-manifold ${\cal M}^4$ into ${\bf R}\times
\Sigma$, with $\Sigma$ an arbitrary 3-manifold
(${\cal M}^4$ has local coordinates $\{x_0,x_1,x_2,x_3\}$), the problem of
finding
all self-dual metrics was reduced to solving one constraint and one
``evolution'' equation on a field of triads $V_i^a$ on $\Sigma$. That is,

\[
 {\rm Div} V_i^a = 0,
\]

\begin{equation}
{\partial V_i^a \over \partial t} = {1\over 2} \epsilon_{ijk}[V_j,V_k]^a,
\label{ajs}
\end{equation}

\noindent
where $i,j,k=1,2,3$, \cite{ajs}. Thus, all self-dual metrics can be described
in terms of the triad just as

\begin{equation}
g^{ab} = ({\rm det \ V})^{-1} [V_i^aV_j^b\delta^{ij} + V_0^aV_0^b].
\label{ajsmetr}
\end{equation}
where $V_0^a$ is the vector field used in the $3+1$ decomposition.

Several authors \cite{grant}\cite{hus}, beginning with the AJS formulation made
contact
with the Pleba\'nski approach to self-dual gravity \cite{pleba}. In
\cite{grant} Grant has
shown that Eqs.(\ref{ajs}) are related in a very close way with the first
heavenly
equation of ref. \cite{pleba}. It
was quickly recognized that the relation was only a Legendre
transformation on a convenient coordinate chart \cite{fin}. Here the
heavenly equation was brought into a Cauchy-Kovalevski evolution form.

On the other hand in \cite{hus} Husain gives a chiral formulation for the
self-dual gravity. He has shown how self-dual gravity can be derived from a
2-dimensional chiral Model which gauge group corresponds to the group of
symplectic diffeomorphisms (area preserving diffeomorphisms of a
2-surface ${\cal N}^2$, SDiff$({\cal N}^2$)).\footnote{As we make only
local considerations we assume the space ${\cal N}^2$ to be a
two-dimensional simply
connected manifold with local coordinates $\{p,q\}$. This space has a
natural local
symplectic structure given by the local area form $\omega = dp\wedge dq$.
The
group SDiff$({\cal N}^2)$ is precisely the group of diffeomorphisms on
${\cal N}^2$ preserving the symplectic structure $\omega$,  i.e. for all $g\in
{\rm SDiff}({\cal N}^2)$, $g^*(\omega) = \omega$.}  Similarly to Grant,
starting from Eqs. (\ref{ajs}) but using another
choice of the set of vector fields $V_i^a$, Husain derived also the
first heavenly
equation. However, although the choice of vector fields is different,
both formulations
are equivalent to that one from Pleba\'nski. Thus, we have a class of
equations (and therefore the corresponding class of solutions) which will be
equivalent. This class of equations we call Grant-Husain-Pleba\'nski GHP
and they can be seen as equivalent to AJS equations. This because they
are only different formulations of the same full theory.

Here, we briefly review the Husain chiral model for self-dual
gravity. It is well known that Equations (\ref{ajs}) lead to the set

\[
[{\cal T},{\cal X}] = [{\cal U},{\cal V}] = 0,
\]

\begin{equation}
[{\cal T},{\cal U}] + [{\cal X},{\cal V}] = 0,
\label{hus}
\end{equation}

\noindent
where ${\cal T} := V_0+iV_1$, ${\cal U} := V_0-iV_1$,
${\cal X} := V_2-iV_3$, ${\cal V}:=V_2+iV_3$. The vector fields ${\cal X}$ and
${\cal T}$ can be fixed to be

\begin{equation}
{\cal T}= {\partial \over \partial \bar z},
\ \ \ \ \  {\cal X}=
{\partial \over \partial z}.
\end{equation}

\noindent
where the $\bar z = x_0 +ix_1$, $z= x_2-ix_3$, $u=x_0-ix_1$ and
$v=x_2+ix_3$. The bar does not stands for complex conjugation. The choice of
vector fields enables four possibilities:

\vskip 1truecm

\noindent
{\it i)} {\it Husain} \cite{hus} (See also \cite{chak}):

Taking

\[
{\cal U} = - \Omega_{,zq}{\partial \over \partial p} +
\Omega_{,zp}{\partial \over \partial q},
\]

\begin{equation}
{\cal V} = \Omega_{, \bar z q}{\partial \over \partial p} -
\Omega_{, \bar z p} {\partial \over \partial q},
\label{huseq}
\end{equation}

\noindent
where $\Omega$ is a holomorphic function of its arguments and $p$,$q$ are
local coordinates on the two-manifold ${\cal N}^2$. Eqs.(\ref{hus}) lead
directly to first heavenly equation as usual \cite{pleba}

\begin{equation}
\Omega_{,zp} \Omega_{,\bar z q} - \Omega_{,zq}\Omega_{,\bar zp} = 1,
\label{pleba}
\end{equation}

\noindent
where $\Omega_{,zp} = {\partial^2 \Omega \over \partial z \partial p}$, etc.

\vskip 1truecm

\noindent
{\it ii)}
{\it Grant} \cite{grant}:

The difference with respect to the Husain's formalism is just the way in which
the vector fields ${\cal U}$ and ${\cal V}$ are chosen. Grant takes

\[
{\cal U} =  {\partial \over \partial \bar z} - h_{,zq}
{\partial \over \partial p} + h_{,zp}{\partial \over \partial q},
\]

\begin{equation}
{\cal V} =  h_{, \bar z q}{\partial \over \partial p} - h_{, \bar z p}
{\partial \over \partial q}.
\label{grant}
\end{equation}

\noindent
Equations (\ref{hus}) lead to the Grant evolution equation, which is of the
Cauchy-Kovalevski form

\begin{equation}
h_{, \bar z \bar z} + h_{,zp}h_{, \bar z q} - h_{,zq}h_{, \bar z p} = 0,
\label{granteq}
\end{equation}

\noindent
where the corresponding metric is

\begin{equation}
g = d\bar z \otimes (h_{,\bar z q} dq + h_{,\bar z p} dp) +
dz \otimes (h_{,zq}dq + h_{,zp} dp) +
{1\over h_{,\bar z \bar z}}(h_{,\bar z q}dq +
h_{,\bar z p}dp)^2.
\label{granmetr}
\end{equation}

After a Legendre transformation on the variable ${\bar z}$ we recover the
first heavenly equation as usual \cite{fin}.

\vskip 1truecm
\noindent
{\it iii)} {\it Grant (a variant)} \cite{grant}:

This choice leads to a formulation similar to that of Grants. Choosing
the vector fields as

\[
{\cal U} = - h_{,zq}{\partial \over \partial p} + h_{,zp}{\partial \over
\partial q},
\]

\begin{equation}
{\cal V} = {\partial \over \partial z} +  h_{,\bar z q}
{\partial \over \partial p} - h_{,\bar z p}{\partial \over \partial q},
\label{grant2}
\end{equation}

\noindent
and using once again equations (\ref{hus}),
one arrives at the Grant evolution equation, which is of the
Cauchy-Ko\-va\-\-levski form

\begin{equation}
h_{,zz} + h_{,\bar z q}h_{,zp} - h_{,\bar z p}h_{,zq} = 0.
\label{gran2teq}
\end{equation}

\noindent
The corresponding metric is of course

\begin{equation}
g = d\bar z \otimes (h_{,\bar z q} dq + h_{,\bar z p} dp) + dz \otimes
(h_{,zq}dq + h_{,zp} dp) + {1\over h_{,zz}}(h_{,zq}dq + h_{,zp}dp)^2.
\label{grant2metr}
\end{equation}

And, as before, the first heavenly equation is recovered after a
Legendre transformation.

\vskip 1truecm
\noindent
{\it iv)} {\it Husain} \cite{hus}:

For the self-dual equations (\ref{ajs}) there exists another possibility for an
appropiate selection of the vector fields. This choice leads to  the chiral
equations, which appear to be non-equivalent to that of the GHP class of
equations.
However they might be related to them.

Introducing now two functions ${\cal B}_1(\bar z,z,p,q)$ and
${\cal B}_2(\bar z,z,p,q)$, the vector fields ${\cal U}$ and ${\cal V}$ can be
written in a completely general form in terms of these functions as

\[
{\cal U} = {\partial \over \partial \bar z} + \alpha^{b}
\partial_b {\cal B}_1,
\]

\begin{equation}
{\cal V} = {\partial \over \partial  z} + \alpha^{b}
\partial_b {\cal B}_2,
\label{uvgen}
\end{equation}

\noindent
where $\alpha^{ab} = ({\partial\over \partial p})^{[a}\otimes
({\partial \over \partial q})^{b]}$.

Using equations (\ref{hus}), the above choice of vector fields leads directly
to the set of equations

\[
{\cal B}_{2,\bar z} - {\cal B}_{1,z} + \{ {\cal B}_1,{\cal B}_2 \} =
{\cal F}_{,\bar z}(\bar z,z) + {\cal G}_{,z} (\bar z,z),
\]

\begin{equation}
{\cal B}_{1,\bar z} + {\cal B}_{2,z} =
{\cal F}_{,z}(\bar z,z) - {\cal G}_{,\bar z}(\bar z,z),
\label{prechi}
\end{equation}

\noindent
for the arbitrary functions ${\cal F}$ and ${\cal G}$. In the above equation
$\{,\}$ means the Poisson bracket in the coordinates $p$ and $q$.

Redefining

\[
A_1(\bar z,z,p,q)= {\cal B}_1 + {\cal G}
\]

\noindent
and

\begin{equation}
A_2(\bar z,z,p,q)= {\cal B}_2 - {\cal F},
\label{b-f}
\end{equation}
(\ref{prechi}) transforms into a two-dimensional chiral model on a two-manifold
${\cal M}^2$
with
local coordinates $\{\bar z,z\}$, having as gauge group the group of
area preserving diffeomorphisms of the two-dimensional manifold ${\cal
N}^2$. This two-dimensional chiral model is

\begin{equation}
F = A_{2, \bar z} - A_{1,z} + \{A_1,A_2\}=0.
\label{chi}
\end{equation}
Vanishing curvature $F=0$ implies that the gauge potentials $A_1$ and $A_2$ are
{\it pure gauge}. Thus, we can write the potentials as

\begin{equation}
A_1 = (\partial_{\bar z} g)g^{-1}, \ \ \ \ \   A_2= (\partial_z g)g^{-1},
\label{a1a2}
\end{equation}
where $g: {\cal M}^2 \times {\cal N}^2 \to {\rm SDiff}({\cal N}^2)$ given
by $g(\bar z,z,p,q) \in {\rm SDiff}({\cal N}^2)$. These potentials satisfy

\begin{equation}
A_{1,\bar z} + A_{2,z} = 0.
\label{a1+a2}
\end{equation}

In this paper we work with the chiral formulation for self-dual gravity
as given by Husain. In Sec. 2, using the formalism of chiral equations
approach to Einstein equations we discuss the chiral equations of the
Husain model as harmonic maps in similar
philosophy of \cite{ma15}\cite{ma29}. In Sec. 3 we find a finite dimensional
subalgebra
of the Lie algebra of SDiff$({\cal N}^2)$, and then we use this
reduction to find solutions. We also find
that the system induced by the Husain formalism is completely integrable
at least for this subalgebra. Finally in Sec. 4 we give our final remarks.

\end{section}

\begin{section}
{Chiral Equations as Harmonic Maps}
In this section we shall outline the method of harmonic maps
for solving the chiral equations. This method consists in applying the
harmonic maps ansatz to the chiral equations. Let us explain it.

First we enunciate the following theorem.

{\bf Theorem.} Let $g \in G$ fulfill the chiral equations. The
submanifold of solutions of the chiral equations
$S\subset G$, is a symmetric manifold (the Riemann tensor of $S$ is
covarariantly null, {\it i.e.} $\nabla{\bf R}_S=0$) with metric
\begin{equation}
{\bf l}_S=tr(dg\ g^{-1}\otimes dg\ g^{-1}),
\label{simmetr}
\end{equation}
where $\otimes$ denotes, the symmetric tensor product.
For the proof see refs. \cite{ma29}\cite{neuge}.

The ansatz
consists in supposing that $g$ can be written in terms of harmonic maps.
Let $V_Q$ be a $Q$-dimensional Riemannian space with an isometry group
$ H \subset
 G = {\rm SDiff}(N^2)$. Suppose that $\{\lambda^i\}$ are
 local coordinates of
$V_Q$. Let $\{\phi_s\},\ \ s=1,..., d={\rm dim} H \leq {\rm dim} G =
\infty,\ \ \
{\rm and }\ \ \  \phi_s=\phi^i_s{\partial\over\partial \lambda^i}$ be a
basis of the Killing vector space of $V_Q$ and $\{\xi^s\}$ the dual
basis of $\{\phi_s\}$. We suppose that
\label{g}
\begin{equation}
g=g(\lambda^i,p,q),\ \ \ \ i=1,...,Q
\label{gchi}
\end{equation}

\noindent
where $\lambda^i(z,\bar z)$ are afinne parameters of the minimal surfaces
of $G$, i.e.

\begin{equation}
\lambda^i_{,z \bar z}+
\Gamma^i_{jk}\lambda^j_{,z}\lambda^k_{,\bar z}=0\ \ \ i,j,k=1,...,Q.
\label{supmin}
\end{equation}

The sdiff$({\cal N}^2)$-valued connection 1-form on the two-manifold
${\cal M}^2$ in the basis $\{d \lambda^i\}$ can be written as (see ref.
\cite{flora})
\begin{equation}
A = a_i(z,\bar z,p,q) d \lambda^i = A_1(z,\bar z,p,q) dz + A_2(
z,\bar z,p,q) d\bar z,
\label{A}
\end{equation}
where $A_1(z,\bar z,p,q) = A_1(\lambda^i,p,q) = a_i(\lambda^i,p,q)
\lambda^i_{,z}$ and
$A_2(z,\bar z,p,q)$ $= A_2(\lambda^i,p,q) = a_i(\lambda^i,p,q)
\lambda^i_{,\bar  z}$.
The functions $a_i(\lambda^i,p,q)$ can be expanded in terms of a basis of
a finite dimensional Lie subalgebra $H$ of sdiff$({\cal N}^2)$,
$\{ \sigma_j \}$, $j=1,2,...,d$; that is
\begin{equation}
a_i(\lambda^i,p,q) = \xi_i^{ s}(\lambda^i) \sigma_s(p,q),
\label{ai}
\end{equation}

\noindent
(for details of this method see refs. \cite{ma15}\cite{ma29}).

\vskip 1truecm
{\bf Theorem}. The potentials $A_1(\lambda^i,p,q)=a_i(\lambda^i,p,q)
\lambda^i_{,\bar z}$ and
$A_2(\lambda^i,p,q) = a_i(\lambda^i,p,q) \lambda^i_{,z}$ are solutions of the
chiral equations (\ref{chi}) and (\ref{a1+a2}).

\vskip 1truecm
{\bf Proof}: Using (\ref{supmin}), equation (\ref{a1+a2}) implies that the
quantities $\xi^s_i(\lambda^i)$
are the components of the Killing  vectors of $V_Q$
\[
A_{1,\bar z} + A_{2,z} = (\xi^s_{i;j}+\xi^s_{j;i})\sigma_s
\lambda^i_{z}\lambda^j_{\bar z}= 0.
\]
where $;$ means covariant derivative in $V_Q$. Equation (\ref{chi})
implies that $\{\sigma_s\}$ are the corresponding
hamiltonian functions of the simplectic form $\omega= dp\wedge dq$ on
${\cal N}^2$, i.e.

\begin{equation}
\{\sigma_s,\sigma_t\}=C^r_{st}(p,q)\sigma_r,
\label{sigpois}
\end{equation}

\noindent
where $C^r_{st}$ are functions of $p$ and $q$ only.
\hfill $\sqcup\!\!\!\!\sqcap $

We shall now use the above approach to Einstein's equations
\cite{ma15}\cite{ma29}, in
order to apply them to self-dual gravity. We show that it is possible to
translate
all relevant tools of the AJS formalism in terms of harmonic maps.

For instance the vector fields ${\cal U}$ and ${\cal V}$ are

\[
{\cal U} = {\partial \over \partial \bar z} + \xi^s_i \lambda^i_{,\bar z}
({\partial \sigma_s \over \partial p}{\partial \over \partial q} -
{\partial \sigma_s \over \partial q}{\partial \over \partial p}),
\]

\begin{equation}
{\cal V} = {\partial \over \partial z} + \xi^s_i \lambda^i_{, z} ({\partial
\sigma_s \over \partial p}{\partial \over \partial q} - {\partial \sigma_s
\over \partial q}{\partial \over \partial p}).
\label{uvhar}
\end{equation}

The vectors on ${\bf R}\times\Sigma^3$ are therefore

\[
V_0 = {\partial \over \partial \bar z} + {1\over 2} \xi^s_i
\lambda^i_{,\bar z}({\partial \sigma_s \over \partial p}
{\partial \over \partial q} - {\partial \sigma_s \over \partial q}
{\partial \over  \partial p}),
\]

\[
V_1 = {i\over 2} \xi^s_i \lambda^i_{,\bar z}
({\partial \sigma_s \over \partial p}{\partial \over \partial q} -
{\partial \sigma_s \over \partial q}{\partial \over \partial p}),
\]

\[
V_2 = {\partial \over \partial z} + {1\over 2} \xi^s_i
\lambda^i_{,z}({\partial \sigma_s \over \partial p}{\partial \over
\partial q} - {\partial \sigma_s \over \partial q}{\partial \over
\partial p}),
\]

\begin{equation}
V_3 =- {i\over 2} \xi^s_i \lambda^i_{,z}({\partial \sigma_s \over
\partial p}{\partial \over \partial q} - {\partial \sigma_s \over
\partial q}{\partial \over \partial p}).
\label{Vhar}
\end{equation}

The self-dual metric (\ref{ajsmetr}) can be expressed in terms of harmonic maps

\begin{equation}
g^{-1} = {4\over \xi^n_i \xi^m_j \lambda^i_{,\bar z} \lambda^j_{,z}
\{\sigma_m,\sigma_n\}} [d\bar z \otimes d\bar z + dz \otimes dz +
\xi^s_k({\partial \sigma_s \over \partial p}dp -
{\partial \sigma_s \over \partial q}dq) \otimes d\lambda^k].
\label{g^-1}
\end{equation}

Here it can be observed that similarly to the metric (3.4) of
ref.\cite{hus},
it also appears a singularity for null Poisson brackets (abelian algebra).

For completeness we can write also the inverse of (\ref{g^-1})

\[
g = \{\xi_k^s(\lambda_{,z}^kd\bar z - \lambda_{,\bar z}^kdz)
-{(\xi_k^s)^2[(\lambda_{,\bar z}^k)^2 +(\lambda_{,z}^k)^2]\over \xi_i^m
\xi_j^n \lambda^i_{,\bar z}\lambda^j_{,z}\{\sigma_m,\sigma_n\}} ({\partial
\sigma_s\over \partial p} dq + {\partial \sigma_s\over \partial q}dp)\}
\]

\begin{equation}
\otimes ({\partial \sigma_s \over \partial p}dq + {\partial \sigma_s
\over \partial q}dp).
\label{g}
\end{equation}

\noindent
 From metric (\ref{g^-1}) and (\ref{g}) it is now clear that $C^r_{mn}$ can not
vanish, thus it is not possible to take an abelian algebra in (\ref{sigpois}).

\end{section}

\vskip 2truecm

\begin{section}
{Two Dimensional Subspaces}
 From metric (\ref{g}) we conclude that it is not possible to take one
dimensional subspaces $V_1$ since all one dimensional Riemannian spaces
contains only abelian groups of motion. We consider a two-dimensional
Riemannian space $V_2$. In \cite{ma29} it was shown that the chiral
equations imply that $V_2$ must be a symmetric space. All two-dimensional
Riemannian space is conformally flat. So, the metric of $V_2$ can be
written as \cite{ma27}

\begin{equation}
i^{*}({\bf l}_S)=ds^2_2={d\lambda d\tau\over(1+k\lambda \tau)^2},
\label{2dmetr}
\end{equation}

\noindent
where $i:V_2 \rightarrow S$.
The symmetry of $V_2$ implies that $k=k(p,q)$ only, {\it i.e.}
$k_{,\lambda}=k_{,\tau}=0$, (since
all two-dimensional symmetric space possesses constant curvature). Thus,
$V_2$
contains a three-dimensional isometry group $H$. Three independent Killing
vectors of $V_2$ are
\[
\xi^1={1\over 2V^2}[(k\tau^2+1)d\lambda+(k\lambda^2+1)d\tau],
\]
\[
\xi^2={1\over V^2}[-\tau d\lambda+\lambda d\tau], \ \ \ \ \ \ \
V=1+k\lambda\tau, \]

\begin{equation}
\xi^3={1\over 2V^2}[(k\tau^2-1)d\lambda+(1-k\lambda^2)d\tau].
\label{kill}
\end{equation}

The three Hamiltonian functions $\sigma_s$ fulfill the algebra
\[
\{\sigma_1,\sigma_2\}=-4k\sigma_3,
\]
\[
\{\sigma_2,\sigma_3\}=\ 4k\sigma_1,
\]
\begin{equation}
\{\sigma_3,\sigma_1\}=\ -4\sigma_2,
\label{alg}
\end{equation}

\noindent
in order to have compatibility with the Killing vectors (\ref{kill}). These
Poisson brackets can be seen as three differential equations for the
three functions $\sigma_s$ and the function $k$, so we can take one of them
arbitrarily and determine the other three ones by integration. Knowing
the functions $\sigma_s$ we can determine the potentials $A_1$ and
$A_2$ by means of the formulae
\[
A_1=\xi_i^s\sigma_s\lambda^i_{,\bar z}, \ \ \ \
A_2=\xi_i^s\sigma_s\lambda^i_{,z}, \] in terms of the harmonic maps
$\lambda^i$. The harmonic map equation (\ref{supmin}) transforms in this case
into
\[
\lambda_{,z\bar z}-
{2k\tau\over 1+k\lambda\tau}\lambda_{,z}\lambda_{,\bar z}=0,
\]
\begin{equation}
\tau_{,z\bar z}-
{2k\lambda\over 1+k\lambda\tau}\tau_{,z}\tau_{,\bar z}=0.
\label{nsupmin}
\end{equation}

In what follows we will solve equations (\ref{alg}). Let us write equation
(\ref{alg}) in terms of two new variables $s=s(p,q)$ and $t=t(p,q)$ and without
loss of generality we can suppose that $\sigma_2=s$. The commutation relations
(\ref{alg}) transform

 into
\[
i)\ \ \ \ \ \ \ \ {\partial\sigma_1\over\partial t}\{s,t\}=4k\sigma_3,
\]
\[
ii)\ \ \ \ \ \ \ \ {\partial\sigma_3\over\partial t}\{s,t\}=4k\sigma_1,
\]
\begin{equation}
iii)\ \ \ \ \ \ \ \
\left({\partial\sigma_3\over\partial s}{\partial\sigma_1\over\partial t}-
 {\partial\sigma_1\over\partial s}{\partial\sigma_3\over\partial t} \right)
  \{s,t\}=-4\sigma_2.
\label{nalg}
\end{equation}

If we substitute ($i)$) and ($ii)$) into ($iii)$) we arrive at

\begin{equation}
k{\partial(\sigma_3^2-\sigma_1^2)\over\partial s}=-2s,
\label{eq1}
\end{equation}

\noindent
and by combining  ($i)$) and ( $ii)$) we conclude that

\begin{equation}
{\partial(\sigma_3^2-\sigma_1^2)\over\partial t}=0,
\label{eq2}
\end{equation}

\noindent
which imply that $k$ does not depend on $t$, that means $k=k(s)$.
Deriving equations ($i)$) and ($ii)$) with respect to $t$ we find differential
equations only for $\sigma_1$ and $\sigma_3$

\begin{equation}
{\partial^2\sigma_s\over\partial t^2}l^2 +
{1\over 2}{\partial\sigma_s\over\partial t}{\partial l^2\over\partial t}-
16k^2\sigma_s=0,
\label{eq3}
\end{equation}
\[
s=1,3
\]
\noindent
where we have define $l=\{s,t\}$. The solution to equation (\ref{eq3}) is

\begin{equation}
\sigma_s=\left[a_s\ e^{L(t)}+b_s\ e^{-L(t)}\right],
\end{equation}
\noindent
where
\[
\noindent
L(t)=4k\int {dt\over l}.
\]
 From (\ref{eq2}) we find that
$a_1=a_2=c_1$ and $b_1=-b_2=c_2$, such that
\[
\sigma_3^2-\sigma_1^2=4\ c_1c_2.
\]
The no dependence on  $t$ of $\sigma_3^2-\sigma_1^2$ implies that
$c_1=c_1(s),\ c_2=c_2(s)$ where $2k{\partial c_1c_2\over\partial s}=-s$.
So we obtain
\[
\sigma_1=\left[ c_1(s)\ e^{L(t)}+c_2(s)\ e^{-L(t)}\right],
\]
\[
\sigma_2=s,
\]
\[
\sigma_3=\left[ c_1(s)\ e^{L(t)}-c_2(s)\ e^{-L(t)}\right],
\]
\begin{equation}
L(t)=4k\int {dt\over l},\ \ \ \ \
2k{\partial c_1c_2\over\partial s}=-s,\ \ \ \ \ \
\{s,t\}=l.
\end{equation}

Observe that $k(s),\ c_1(s)$ and $c_2(s)$ are subjected to only one
restriction, therefore two of them are arbitrary. So we have three
arbitrary functions of $p,q$ in general. \end{section}

\begin{section}
{Final Remarks}

In this paper we found an explicit exact class of solutions to self-dual
gravity \cite{hus}. We used the chiral equations approach in
order to obtain explicit solutions. Solving the chiral equations with
the harmonic maps method we find that the harmonic maps ansatz can be
applied to the chiral equations  derived from self-dual gravity. The
difference with previous applications of this method is that here we have
Poisson brackets in place of matrix brackets in a similar spirit as in
\cite{flora}. Nevertheless we can solve the corresponding Poisson
algebra by
making a coordinate transformation and finding the corresponding
hamiltonian functions by solving the Poisson algebra as differential
equations. We find that there exists a class of such solutions in terms
of two arbitrary functions ($s$ and $t$) of two variables ($p$ and $q$).
The coordinate transformation can be taken
also arbitrary, but in the case when the new coordinates are canonical
the solution becomes very simple.

\end{section}
\vskip 2truecm
\begin{center}
{\bf Acknowledgements}
\end{center}
We are indebted to A. Ashtekar, R. Capovilla, J.D.E. Grant, L.E.
Morales, J.F. Pleba\'nski and the Referee for useful comments and
suggestions. This work is partially supported by  CONACyT-M\'exico.

\end{document}